\documentclass[prb,aps,reprint,twocolumn,superscriptaddress,floatfix,showpacs]{revtex4-2}

\usepackage{graphicx}
\usepackage{amssymb}
\usepackage{amsmath}
\usepackage{color}
\usepackage{changes}
\usepackage{hyperref}
\usepackage{comment}
\usepackage{lipsum}
\usepackage{paralist}
\usepackage{ dsfont }   			    
\usepackage{float}					
\usepackage[caption=false,position=top,subrefformat=parens]{subfig}

\usepackage{notes2bib}
\bibnotesetup{
	note-name = ,
	use-sort-key = false
}

\newcommand{\timerev}[1]{} 


\begin{document}

\title{Transition between scattering regimes of 2D electron transport}
\author{Philipp Heilmann}
\email[Email:\,]{philipp.heilmann@uni-wuerzburg.de}
\affiliation{ Institute of Theoretical Physics and Astrophysics, University of Würzburg, Germany}  
\author{Pavlo~V.~Pyshkin}
\affiliation{Physikalisches Institut, Experimentelle Physik III, Universit\"at W\"urzburg, Am Hubland, 97074 W\"urzburg, Germany}
\affiliation{Institute for Topological Insulators, Am Hubland, 97074 W\"urzburg, Germany}
\author{Björn Trauzettel}
\affiliation{ Institute of Theoretical Physics and Astrophysics, University of Würzburg, Germany} 
\affiliation{Würzburg-Dresden Cluster of Excellence ct.qmat, Germany} 

\date{\today}

\begin{abstract}
We examine 2D electron transport through a long narrow channel driven by an external electric field in presence of diffusive boundary scattering. At zero temperature, we derive an analytical solution of the transition from ballistic to diffusive transport if we increase the bulk disorder strength. This crossover yields characteristic current density profiles. Furthermore, we illustrate the current density in the transition from ballistic to hydrodynamic transport. This corresponds to the Gurzhi effect in the resistivity. We also study the influence of finite temperature on current densities and average current in this system. In particular, we analyze how a particular scaling law of electron-electron scattering with respect to temperature affects the current along the channel.
\end{abstract}
\maketitle

\section{Introduction}
In recent years, the field of two dimensional (2D) electron transport has advanced remarkably \cite{Kukushkin2004,Weber2005,Spivak2006,Bykov2007,Khodas2010,Dai2010}. 
The emergence of ultraclean two dimensional materials like Ga(Al)As heterostructures, graphene or transition-metal dichalcogenides (TMDCs) has sparked developments both experimentally and theoretically \cite{Mueller2008, Bockhorn2011, Lee2011, Hatke2012, Mani2013, Yigen2013, Mendoza2013, Shi2014, Narozhny2015, Moll2016, Ledwith2017, Lucas2017, Alekseev2018, Gooth2018, Sulpizio2019, Alekseev2019, Holder2019, Raichev2020, Afanasiev2021, Steinberg2022, Kryhin2023, Hofmann2023, Zohrabyan2023, Kryhin2023a, Kryhin2023b}.
In 2D systems, bulk scattering is less dominant than in their 3D counterparts. Consequently, it is possible to efficiently reduce impurity scattering in these systems. This possibility allows us to explore other than diffusive transport regimes in the laboratory, such as ballistic transport, hydrodynamic transport and their crossovers.
Special focus lies on channel geometries as a setup for a 2D system with boundaries \cite{Sulpizio2019, Alekseev2018, Alekseev2019, Afanasiev2021, Holder2019, Raichev2020}. 
Particular transport regimes show characteristic flow profiles, which are observable in the laboratory \cite{Sulpizio2019, Ella2019}. \\
A theory of ballistic transport of a 2D electron gas in presence of electron-electron interactions in a channel has been developed in Ref. \cite{Alekseev2018}.
It has been predicted that the average current diverges logarithmically for weak electron-electron scattering. This divergence stems from longitudinal modes that do not scatter at the boundaries of the channel. 
We further develop this theory by different means: \begin{inparaenum}[(i)]
\item We derive an analytical solution based on Meijer-G functions for the current density and the average current valid for arbitrary impurity scattering and weak electron-electron scattering rates.
\item We study the current density if we crossover between different transport regimes (ballistic, diffusive, hybrodynamic). Interestingly, the profile of the current density for ballistic transport and hydrodynamic transport (with diffusive boundaries in a channel geometry) look alike. Those profiles differ, however, for diffusive transport.
\item We analyze the temperature dependence of transport for particular scaling laws of 2D electron-electron scattering. We obtain analytical results under the assumption that impurity scattering is dominating over electron-electron scattering.
\end{inparaenum} \\
This article is organized as follows. We introduce our model of a 2D electron system in a channel geometry in Sec. \ref{modelsec}. 
In Sec. \ref{ballistictransport}, we present the analytic solution for the current density and the average current describing the transition from ballistic to diffusive transport.
In Sec. \ref{currentdensityprofiles}, we show current density profiles illustrating the transitions between different transport regimes (ballistic, diffusive, hydrodynamic).
In Sec. \ref{nonzerot}, we extend the theory to finite temperature. We pay particular attention to the temperature dependence of the average current in channel geometry. We conclude in Sec. \ref{conclusion}.
\vspace{-0.4cm}
\section{Model}
\label{modelsec}
\begin{figure}[t]
    \centering
    \includegraphics[width=0.5\columnwidth]{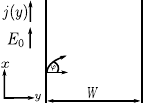}
    \caption{Schematic illustration of the 2D electron channel of width $W$ with an external electric field $E_0$ applied in longitudinal direction. }
    \label{model}
\end{figure}
 We consider 2D electron transport through a long but narrow sample of length $L$ and width $W$, where $L\gg W$, schematically illustrated in Fig. \ref{model}.
 We are interested in linear response to a homogeneous electric field $E_0$ along the sample in x-direction.
 The response is described by a non-equilibrium part in the distribution function $f$. Hence, the distribution function takes the form
\begin{equation}
	f = f_0 + \delta f,
	\label{distfunc}
\end{equation} 
where $f_0$ is the equilibrium Fermi distribution.\\
The distribution function depends on a number of parameters related to the setup, the applied field and the scattering mechanisms that we analyze in our work.\\
For the non-equilibrium part of the distribution function $\delta f$, it is common to use the ansatz
\begin{equation}
	\delta f = \left(-\frac{\partial f_0}{\partial \epsilon}\right)h,
	\label{energydep}
\end{equation}
where $\epsilon$ is the electron energy. 
In semiclassics, which is the regime we are interested in, $f$ is a function of time, space and momentum $f(\vec{x},\vec{p},t)$. 
We directly drop the time-dependence as we want to study stationary solutions. 
Since we assume $L\gg W$, we also drop the $x$ dependence as our distribution should not depend on the longitudinal direction in this regime.
This leaves us with $f=f(y,\vec{p})$. 
Next we introduce the angle $\varphi$, chosen as the angle of the electron velocity with the normal of the left sample edge, \textit{cf.} Fig. \ref{model}.
The electron velocity and therefore also the electron momentum is expressed through this angle via 
\begin{equation}
	\vec{p} = m\cdot\vec{v} = m\cdot v(\epsilon)\cdot \left(\begin{array}{c}\sin(\varphi)\\\cos(\varphi)\end{array}\right),
	\label{momentum}
\end{equation}
where $m$ is the electron mass. The corresponding kinetic equation is given by
\begin{equation}
	v(\epsilon) \left(\cos(\varphi)\frac{\partial h}{\partial y} - \sin(\varphi)eE_0\right) = St[h],
	\label{kineq1}
\end{equation}
The collision integral $St[h]$ on the right handside of Eq. $\eqref{kineq1}$ includes electron-electron scattering and scattering due to disorder.
We use a simplified form of the collision integral that satisfies these constraints \cite{Holder2019, Lucas2017}:
\begin{equation}
	St[h] = -\gamma h + \gamma_{ee}\hat{P}[h] + \gamma_{d} \hat{P}_0[h],
	\label{collint}
\end{equation}
where $\gamma = \gamma_{d} + \gamma_{ee}$ is the total (energy-independent) scattering rate; $\gamma_{ee}$ and $\gamma_{d}$ are the electron-electron and disorder scattering rates, respectively. 
$\hat{P}_0$ and $\hat{P} = \hat{P}_{-1} +\hat{P}_0 + \hat{P}_{1}$ are projectors of $h(\varphi)$ onto the subspaces $\{1\}$ and $\{1, e^{\pm i\varphi}\}$. 
Since we consider zero magnetic field the terms proportional to $\hat{P}_0[h]$ vanish \cite{Holder2019,Raichev2020}, which implies particle conservation.
In the analytical results presented in Secs.~III and V below, we neglect the $\hat{P}[h]$ term, which is a reasonable approximation if $\gamma_{ee}\ll \gamma_d$ \cite{Alekseev2018}.
However, we keep the $\hat{P}[h]$ term in our numerical analysis presented in Sec.~IV.\\
We assume fully diffusive Fuchs boundary conditions~\cite{Fuchs_1938} \footnote{However it is straightforward to generalize the formalism to a mix of diffusive and specular boundary conditions \cite{Holder2019}}, parametrized as
\begin{align}
	h(-W/2,\varphi) &= c_l \;\;\;\;\; -\frac{\pi}{2} < \varphi < \frac{\pi}{2},\\
	h(W/2,\varphi) &= c_r \;\;\;\;\;\;\; \frac{\pi}{2} < \varphi < \frac{3\pi}{2}.
	\label{bc}
\end{align}
Here, $c_l$ and $c_r$ are quantities that correspond to the average value of the distribution function over all angles associated with electrons reflected from the respective other edge.
They are defined as
\begin{align}
	c_l &= -\frac{1}{2}\int_{\pi/2}^{3\pi/2} d\varphi' \cos(\varphi') h(-W/2,\varphi'),\\
	c_r &= \frac{1}{2}\int_{-\pi/2}^{\pi/2} d\varphi' \cos(\varphi') h(W/2,\varphi').
\end{align}
Due to these boundary conditions, the probability of an electron being reflected on the sample edge is independent of the reflection angle $\varphi$ at a given boundary.
Moreover, the transverse component of the electron flow vanishes at the edges and therefore, due to the continuity equation with the absence of magnetic field, everywhere.
These considerations imply that $c_l + c_r = 0$. For any antisymmetric function $h(\pm W/2,\varphi)$ in $\varphi$ this results in $c_l=c_r=0$.\\
Combining Eqs. $\eqref{kineq1}$ and $\eqref{collint}$ with the aforementioned approximation, we obtain the simplified kinetic equation
\begin{equation}
	\cos(\varphi)\frac{\partial h}{\partial y} - eE_0\sin(\varphi) = -\gamma' h
	\label{reference1}
\end{equation}
with $\gamma'=\frac{\gamma}{v(\epsilon_F)}$.
In Eq. \eqref{reference1}, we replace $\epsilon$ by $\epsilon_F$ in the velocity $v(\epsilon)$, as we are interested in the low temperature regime, $k_BT \ll \epsilon_F$, for now.
All terms in Eq. \eqref{reference1} have the dimension [energy/length]. We divide the equation by $\left( \frac{\epsilon_F}{W}\right)$ to arrive at a dimensionless equation. Introducing the dimensionless variables 
\begin{align}
	\hat{h}&=\frac{h}{\epsilon_F} ,          &  \hat{E}_0 &=\frac{eE_0W}{\epsilon_F},    \nonumber          \\
	\hat{y}&=\frac{y}{W} ,        &  \hat{\gamma} &= \gamma'W,     \nonumber	     
	\label{dimlessvariables} 
\end{align}
and inserting them into the kinetic equation, this yields
\begin{equation}
	\cos(\varphi)\frac{\partial \hat{h}}{\partial \hat{y}} - \hat{E}_0\sin(\varphi) = -\hat{\gamma} \hat{h} \ ,
	\label{kineqdimless2}
\end{equation}
which we now analyze in detail.
\section{Crossover from ballistic to diffusive transport}
\label{ballistictransport}
    \begin{figure*}[t!]
            \vspace{-0.5cm}
            \subfloat[\phantom{testtestetstetstesttesttesttesttesttesttesttesttest}]{%
            	\includegraphics[width=.48\linewidth]{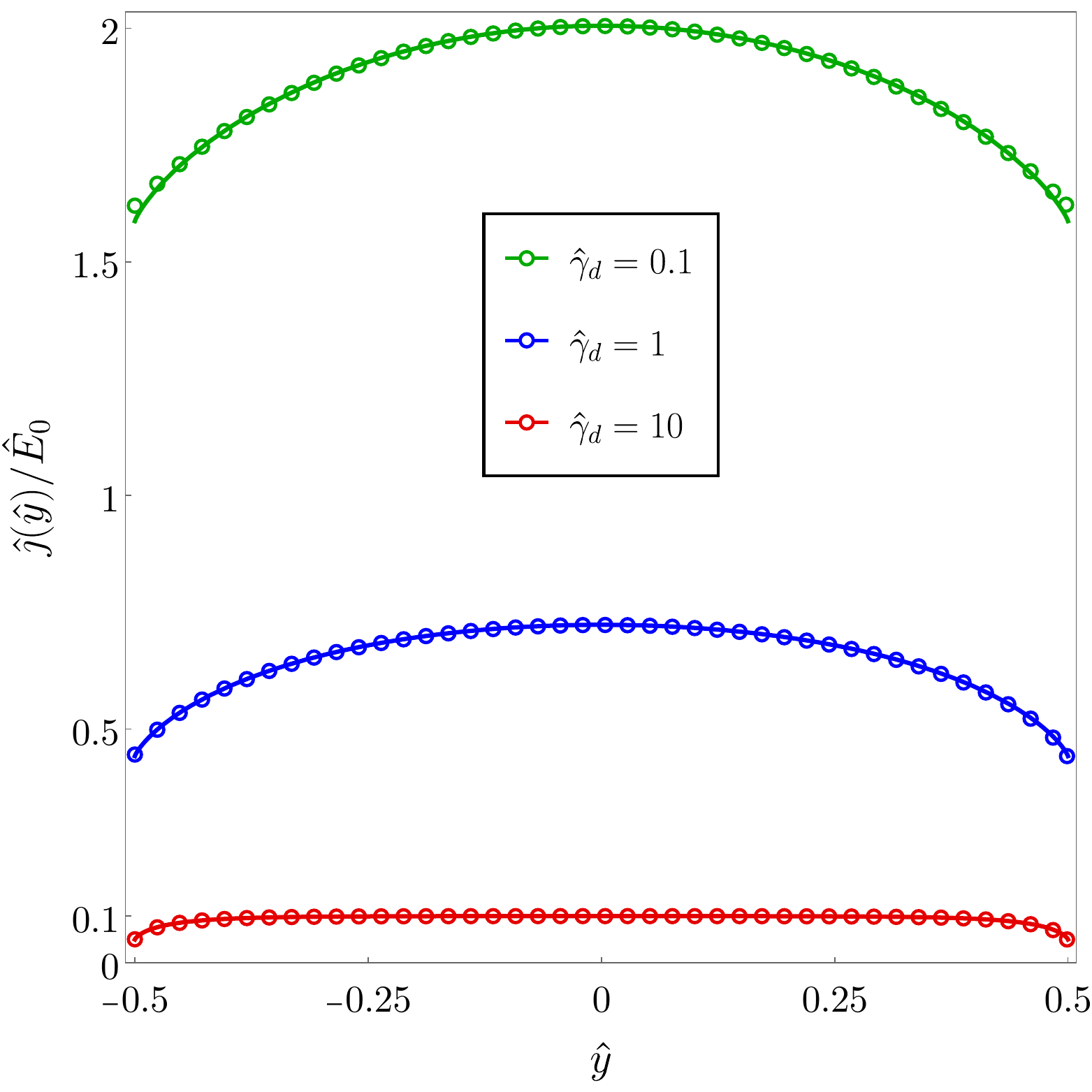}%
            	\label{subfig:a}%
            }\hfill
            \vspace{-0.5cm}
            \subfloat[\phantom{testtestetstetstesttesttesttesttesttesttesttesttest}]{%
            	\includegraphics[width=.48\linewidth]{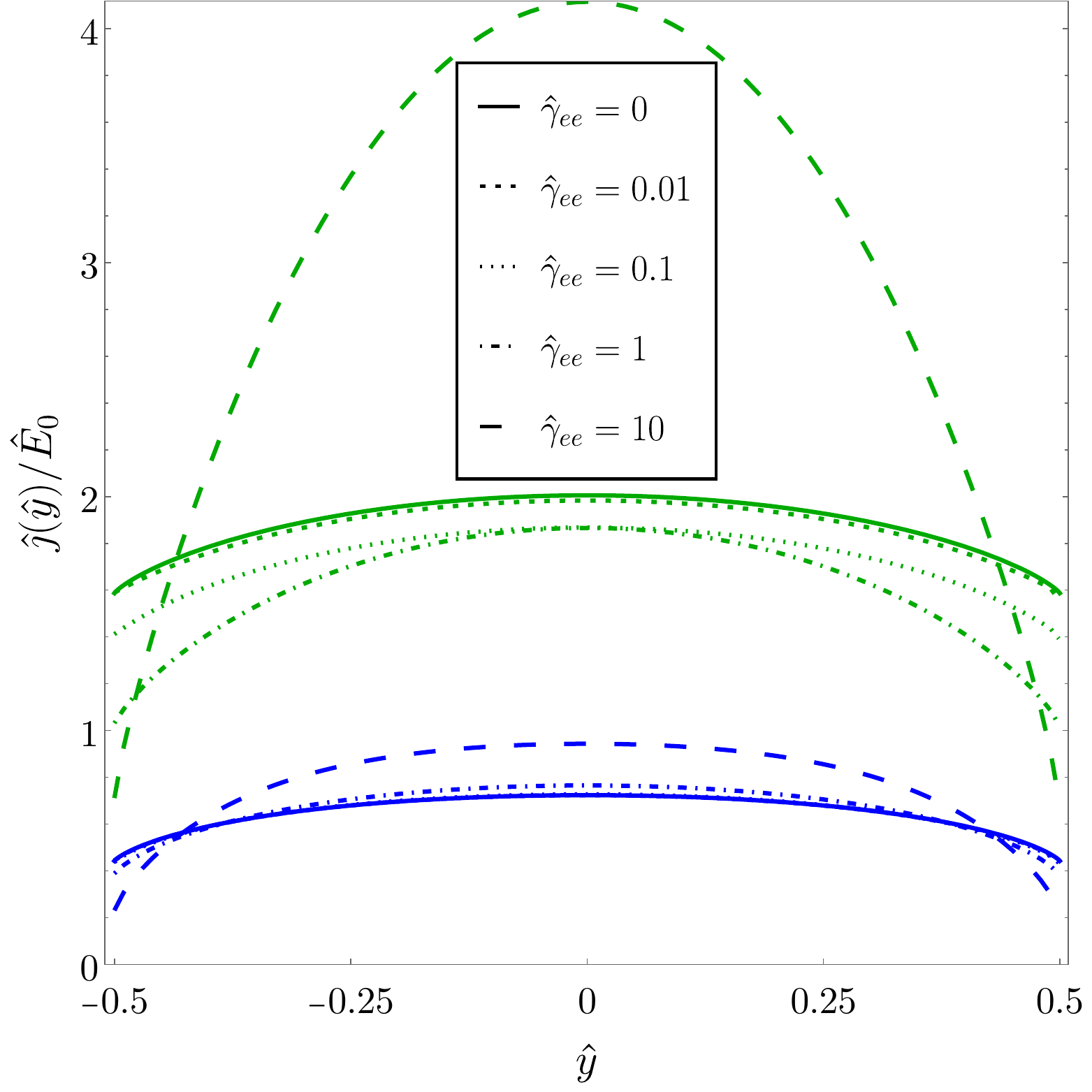}%
            	\label{subfig:b}%
            }\\
            \vspace{-0.2cm}
            \subfloat[\phantom{testtestetstetstesttesttesttesttesttesttesttesttest}]{%
            	\includegraphics[width=.48\linewidth]{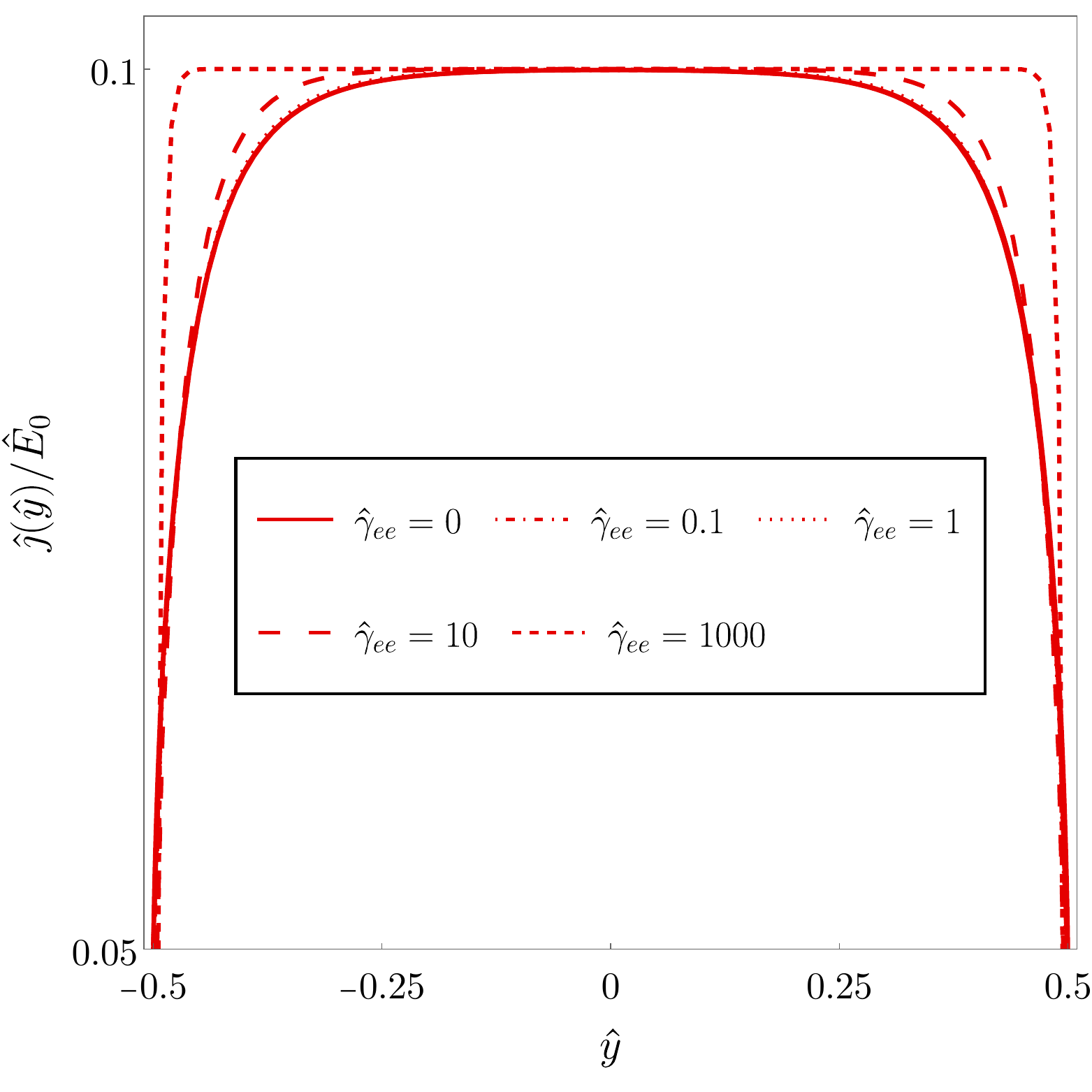}%
            	\label{subfig:c}%
            }\hfill
            \vspace{-0.2cm}
            \subfloat[\phantom{testtestetstetstesttesttesttesttesttesttesttestte}]{%
            	\includegraphics[width=.48\linewidth]{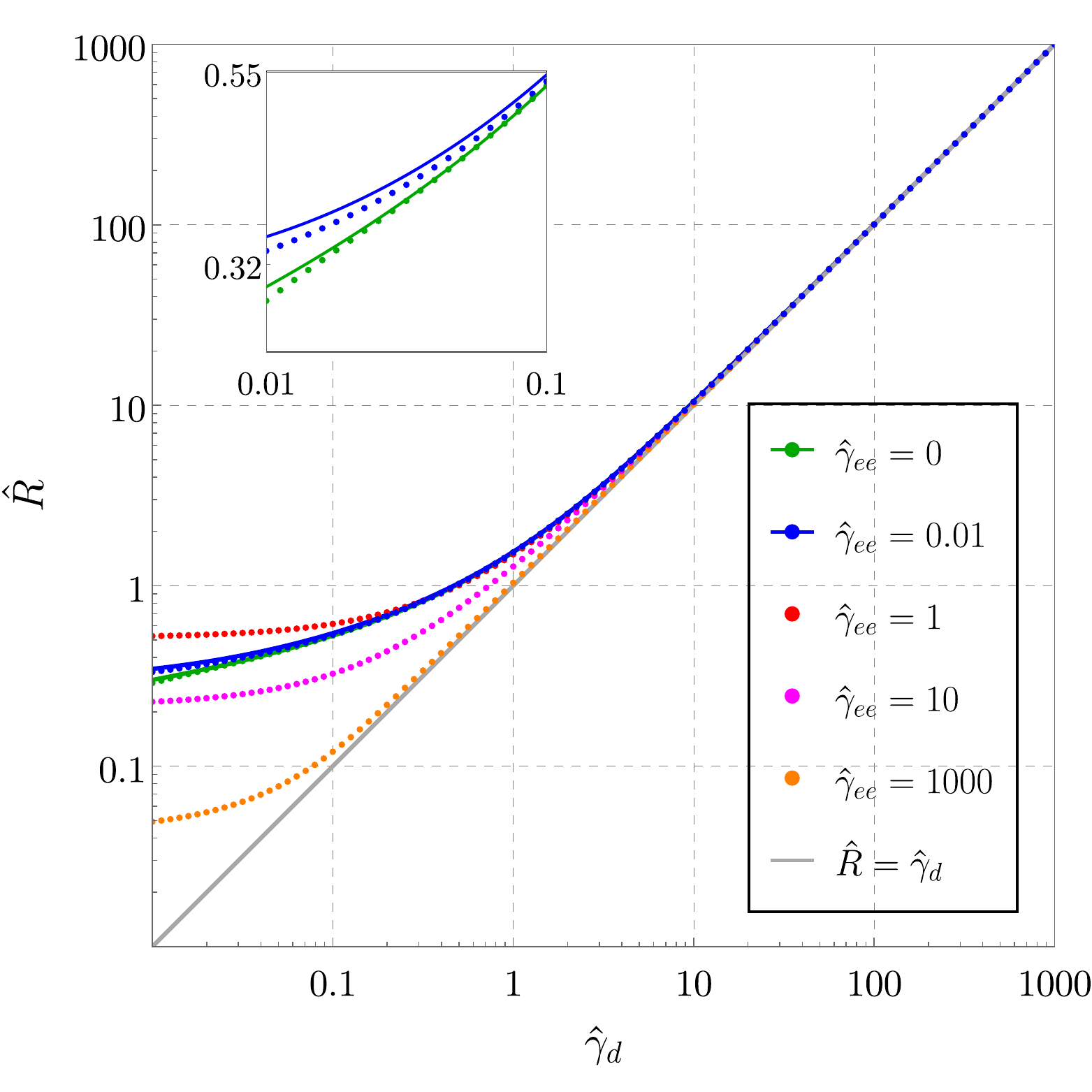}%
            	\label{subfig:d}%
            }
        \caption{(a) Transition from ballistic ($\hat{\gamma}_d=0.1$) to diffusive ($\hat{\gamma}_d=10$) transport in the current density $\hat{\jmath}(\hat{y})$. Dots are numerical results, full lines are corresponding analytical results. (b) Transition from ballistic to hydrodynamic transport as a function of $\hat{\gamma}_{ee}$ in the current density $\hat{\jmath}(\hat{y})$ for $\hat{\gamma}_d=0.1$ (green curves) and $\hat{\gamma}_d=1$ (blue curves). (c) Current density $\hat{\jmath}(\hat{y})$ for $\hat{\gamma}_d=10$ at different values of $\hat{\gamma}_{ee}$. (d) Resistance $\hat{R}=\frac{\hat{E}_0 W}{\hat{I}}$ as a function of $\hat{\gamma}_d$ at different values of $\hat{\gamma}_{ee}$. The inset zooms in and compares the numerical result (dots) and the analytical result (solid line) given by Eq. \eqref{currentsol} in the regime $\hat{\gamma}_{ee} \ll \hat{\gamma}_{d}$. In panels (b), (c), and (d), full lines show analytical results and all other lines (dashed, dotted, dashed-dotted) are numerical results.}
        \label{plots}
    \end{figure*}
It is straightforward to solve Eq. \eqref{kineqdimless2} with boundary conditions $\eqref{bc}$ \cite{Alekseev2018}. The result can be written as
\begin{align}
	\hat{h}_{\pm}(\hat{y},\varphi) = \frac{\hat{E}_0 \sin(\varphi)}{\hat{\gamma}}\left(1- \exp\left(-\frac{\hat{\gamma} (\hat{y}\pm \frac{1}{2})}{\cos(\varphi)}\right)\right),
	\label{solnomf2}
\end{align}
where the index $\pm$ indicates the solution satisfying the boundary condition for the left and right edge respectively.
We are interested in the current density in longitudinal direction
\begin{equation}
	j_x(y) = 2e \int \,v_x\, \delta f \frac{d\vec{p}}{(2\pi \hbar)^2},
	\label{current_density_x}
\end{equation}
and the corresponding average current $I=\int dy j_x(y)$. For simplicity, we normalize the current density and drop the index $x$ to arrive at $\hat{\jmath}(\hat{y})\!=\!\frac{j_x(\hat{y})}{e \mathcal{D}v(\epsilon_F)}$, where $\mathcal{D}=\frac{m}{\pi \hbar^2}$ is the two dimensional density of states. Then, the normalized, longitudinal current density can be expressed as 
\begin{align}
	\hat{\jmath}(\hat{y}) &= \int_{-\pi/2}^{\pi/2} \frac{\hat{E}_0 \sin(\varphi)^2}{\pi \hat{\gamma}}\left(1- \exp\left(-\frac{\hat{\gamma} (\hat{y}+ \frac{1}{2})}{\cos(\varphi)}\right)\right) d\varphi \label{currentdensity} \\
	&+\int_{\pi/2}^{3\pi/2}  \frac{\hat{E}_0 \sin(\varphi)^2}{\pi \hat{\gamma}}\left(1- \exp\left(-\frac{\hat{\gamma} (\hat{y}- \frac{1}{2})}{\cos(\varphi)}\right)\right) d\varphi \; .\nonumber
\end{align}
The analytical solution of the remaining integral over the angle $\varphi$ is given by
\begin{align}
	\hat{\jmath}(\hat{y}) &= \frac{\hat{E}_0}{8\hat{\gamma}}   \left(\sum_{\pm}(1\pm 2\hat{y})^2\hat{\gamma}^2 \phantom{\frac{1}{2}}\right. \label{currentdensitysol} \\ 
	&-\left.4 \pi  G_{2,4}^{2,0}\left(\frac{1}{16} (1\pm 2\hat{y})^2\hat{\gamma}^2|
	\begin{array}{c}
		0,2 \\
		\frac{1}{2},\frac{1}{2},0,0 \\
	\end{array}
	\right)\right), \nonumber
\end{align}
where $G_{2,4}^{2,0}\left(\frac{1}{16} (1\pm 2\hat{y})^2\hat{\gamma}^2|
\begin{array}{c}
	0,2 \\
	\frac{1}{2},\frac{1}{2},0,0 \\
\end{array}
\right)$ denotes the Meijer-G function. The average current associated with this current density takes the form
\begin{equation}
	\hat{I} = \frac{\hat{E}_0 W\hat{\gamma}}{3}-\frac{\pi \hat{E}_0 W}{2 \hat{\gamma} } G_{3,5}^{2,1}\left(\frac{\hat{\gamma}^2}{4}|
	\begin{array}{c}
		\frac{1}{2},0,2 \\
		\frac{1}{2},\frac{1}{2},-\frac{1}{2},0,0 \\
	\end{array}
	\right).
	\label{currentsol}
\end{equation}
\\
This result for the current is valid for arbitrary $\hat{\gamma}$ within the validity regime of the Boltzmann equation \eqref{reference1}. In the limit $\hat{\gamma}\rightarrow0$, we obtain in leading order
\begin{equation}
    \frac{\hat{I}}{\hat{E}_0}=\frac{2 W}{\pi} \left(\ln \left(\frac{1}{\hat{\gamma}} \right)+\ln (2)+\frac{1}{2}- \varGamma\right),
    \label{molenkampresult}
\end{equation}
where $\varGamma$ is Euler's constant. Note that Eq. \eqref{molenkampresult} resembles Eq.~(22) in Ref. \cite{Jong1995}.
For $\hat{\gamma} \gg 1$, the current given by Eq. \eqref{currentsol} becomes $\hat{I} \propto \frac{1}{\hat{\gamma}}$.
This is visualized via the resistance $\hat{R}=\frac{\hat{E}_0}{\hat{I}}\propto \hat{\gamma}$ in Fig. \subref*{subfig:d}. It marks the Drude regime, where boundary scattering is not contributing to the current. 

We emphasize that Eqs. \eqref{currentdensitysol} and \eqref{currentsol} describe analytically the full crossover from ballistic to diffusive transport in a channel geometry with diffusive boundary conditions.

\section{Current density profiles}
\label{currentdensityprofiles}
In the following, we present and analyze current density profiles for different transport regimes (ballistic, diffusive, hydrodynamic). Some results are based on analytical solutions presented in the previous section. Others are derived numerically.\\
In the limit $\hat{\gamma}\ll1$, the current density \eqref{currentdensitysol} shows parabolic behaviour, illustrated in Fig. \subref*{subfig:a}. This feature is somewhat counter intuitive but known. It can be viewed as a current density that resembles the shape of hydrodynamic electron flow in weakly interacting systems (hydro without hydro). It stems from the combination of ballistic transport with diffusive boundary scattering. For $\hat{\gamma}>1$, we observe the transition to the diffusive regime, which has the characteristic flat form of the current density \cite{Zohrabyan2023} due to dominant bulk-impurity scattering. This crossover from ballistic to diffusive transport is fully described by our analytical results. \\
Note that Eqs. \eqref{currentdensitysol} and \eqref{currentsol} also describe transport regimes, in which electron-electron scattering is relevant as long as the corresponding scattering rate is weak enough, see Fig. \subref*{subfig:b}. Exploiting the numerical methods described in Ref. \cite{Huang2021}, we solve Eq. \eqref{kineq1} taking into account all terms in the collision integral \eqref{collint} with two parameters $\hat{\gamma}_{ee}$ and $\hat{\gamma}_d$. Then, we are able to reproduce the results of Ref. \cite{Jong1995}. However, we go beyond these results by a careful comparison of the current densities if we crossover between different scattering regimes. We mention in passing that the numerical results for $\hat{\gamma}_{ee}=0$ coincide with the analytical results given by Eqs. \eqref{currentdensitysol} and \eqref{currentsol} \footnote{For small values of $\hat{\gamma}_{d}$, the numerical result for the current density close to the boundaries differ slightly from the analytical result. This is a finite size effect.}. \\
In Fig. \subref*{subfig:b}, we show the transition from ballistic to hydrodynamic transport for two different values of $\hat{\gamma}_{d}$. We find that the transition to the hydrodynamic regime, characterized by the parabolic form of the current density, occurs only for smaller values of $\hat{\gamma}_{d}$. Instead, for larger values of $\hat{\gamma_d}$, impurity scattering in the bulk is dominant. This is most pronounced for $\hat{\gamma}_{d}\gg 1$, as illustrated in Fig. \subref*{subfig:c}. In this case, the current density can not take a parabolic form anymore even for very large values of $\hat{\gamma}_{ee} \gg \hat{\gamma}_{d}$. Increasing the electron-electron scattering rate then just enhances the flat range of the current density. This phenomenon can also be observed in the resistance, which becomes linear for $\hat{\gamma}_{d}\gg1$ for all values of $\hat{\gamma}_{ee}$, displayed in Fig. \subref*{subfig:d}. The linear behavior sets in earlier for stronger electron-electron scattering. This occurs because the strong scattering between electrons as well as electron-impurity scattering hinders the majority of electrons from undergoing diffusive scattering at the boundaries. Indeed, the characteristic time for diffusive transition across the channel is $\tau_{b} \approx W^2 v_F^{-2}(\gamma_d + \gamma_{ee})$. Hence, the role of the boundaries becomes negligible in the limit $\tau_b \rightarrow \infty$, which can be achieved by tuning $\gamma_{ee}$ as well as $\gamma_{d}$.\\
In Fig. \subref*{subfig:d}, we illustrate non-monotonic variations of the resistance $\hat{R}$ for fixed values of $\hat{\gamma}_d$. For instance, for small values $\hat{\gamma}_d$, $\hat{R}$ first increases with increasing $\hat{\gamma}_{ee}$ (Knudsen effect) and then decreases with increasing $\hat{\gamma}_{ee}$ (Poiseuille effect). The crossover from Knudsen to Poiseuille as a function of $\hat{\gamma}_{ee}$ is known as the Gurzhi effect. 
 
\section{Finite-temperature current affected by electron-electron scattering}\label{nonzerot}
In this section, we investigate the influence of finite temperature on average current and current density. We restrict ourselves to weak electron-electron scattering, i.e. $\gamma_{ee} \ll \gamma_{d}$. This assumption allows us to neglect the $P[\hat{h}]$ term in Eq. \eqref{collint}. We start with the solution to the kinetic equation displayed in Eq. \eqref{solnomf2} and reintroduce the energy dependence by $v(\epsilon)= \frac{\sqrt{2m\epsilon}}{m}$. This implies that we assume a quadratic dispersion relation of the electrons. In this section, we keep the physical dimension of the relevant parameters and observables of our model for now, to precisely consider the energy and temperature dependence of transport. Then, Eq. \eqref{solnomf2} is modified as 
\begin{align}
	\tilde{h}_{\pm}(\tilde{y},\varphi, \epsilon)\! =\! \frac{\tilde{E}_0 \sin(\varphi)\sqrt{2m\epsilon}}{\tilde{\gamma}\, m}\!\!\left(\!1- \exp\!\left(\!-\frac{\tilde{\gamma} (\tilde{y}\pm \frac{1}{2})m}{\cos(\varphi)\sqrt{2m\epsilon}}\right)\!\!\right)\!,
	\label{solnomfdimfull}
\end{align}
where $\tilde{E}_0=eE_0W$, $\tilde{\gamma}=\gamma W$ and $\tilde{y}=\frac{y}{W}$.
The relevant integral for the current density is given by 
\begin{equation}
	\vec{j}(y) = 2e \int \,\vec{v}\, \delta f \frac{d\vec{p}}{(2\pi \hbar)^2},
	\label{generalcurrent}
\end{equation}
where $e$ is the elementary charge, $\vec{v}$ the velocity vector and $\delta f$ the nonequilibrium part of the distribution function. We use the same ansatz, $\delta f = \left(-\frac{\partial f_0}{\partial \epsilon}\right)\, \tilde{h}(\tilde{y},\varphi, \epsilon)$, as before. Next,  we transform the momentum vector $\vec{p}$ to polar coordinates and switch from momentum to energy integration. This results in
\begin{align}
	\vec{j}(\tilde{y}) = \int d\epsilon \int d\varphi \frac{e\sqrt{2m\epsilon}}{2\pi^2 \hbar^2} \left(-\frac{\partial f_0}{\partial \epsilon}\right) \left(\begin{array}{c}\sin(\varphi)\\\cos(\varphi)\end{array}\right)  \tilde{h}(\tilde{y},\varphi,\epsilon).
\end{align}
We now address the current density that stems from charge carriers in an energy window around the Fermi energy $\epsilon_F$ of size $k_BT$, i.e. 
\begin{equation}
	\vec{j}(\tilde{y}) = \frac{e}{2\pi^2 \hbar^2}\int\limits_{\epsilon_F-\frac{k_BT}{2}}^{\epsilon_F+\frac{k_BT}{2}}\!\!\!\!\! d\epsilon \int\limits_{-\frac{\pi}{2}}^{{\frac{\pi}{2}}} d\varphi \frac{\sqrt{2m\epsilon}}{k_B T} \left(\begin{array}{c}\sin(\varphi)\\\cos(\varphi)\end{array}\right) \tilde{h}(\tilde{y},\varphi,\epsilon).
\end{equation}
The prefactor of $1/k_BT$ ensures that, in the limit $T\rightarrow0$, the results of the previous section are recovered. Since the transverse direction of the current density vanishes, we are left with the longitudinal current density $j_x(\tilde{y})$. For simplicity, we drop the index $x$ in the following. Inserting $\tilde{h}_{+}(\tilde{y},\varphi, \epsilon)$ specified in Eq. \eqref{solnomfdimfull}, we obtain for $j_{+}(\tilde{y})$:
\begin{align}
	j_{+}(\tilde{y}) &= \frac{e \tilde{E}_0}{\pi^2 \hbar^2k_B T}\int\limits_{\epsilon_F-k_BT/2}^{\epsilon_F+k_BT/2}\!\! d\epsilon \int\limits_{-\frac{\pi}{2}}^{{\frac{\pi}{2}}} d\varphi \frac{\epsilon}{\gamma}\sin(\varphi)^2\\
	&\left(1- \exp\left(-\frac{\gamma (\tilde{y}+ \frac{1}{2})m}{\cos(\varphi)\sqrt{2m\epsilon}}\right)\right). \nonumber
	\label{currentdensitydefepsilon}
\end{align}
A similar expression can be written down for $j_{-}(\tilde{y})$ determined by $\tilde{h}_{-}(\tilde{y},\varphi, \epsilon)$ specified in Eq. \eqref{solnomfdimfull}.  Solving these integrals, the full current density $j(\tilde{y})= j_{+}(\tilde{y}) + j_{-}(\tilde{y})$ can again be expressed in terms of Meijer-G functions as
\begin{align}
&j(\tilde{y})= \frac{e\tilde{E}_0 m \tilde{\gamma}}{\pi \hbar^2} \left(\frac{1}{4}+\tilde{y}^2\right)+\frac{e\tilde{E}_0 m^2 \tilde{\gamma}^3}{64 \hbar^2 k_B T}\cdot \\
& \left[\left(\frac{1}{2}- \tilde{y}\right)^4 \left(G_{2,4}^{2,0}\left(-\frac{m\tilde{\gamma}^2 (\frac{1}{2}-\tilde{y})^2}{4 (k_B T-2 \epsilon_F)}| \nonumber
\begin{array}{c}
	-2,1 \\
	-\frac{3}{2},-\frac{3}{2},-2,-2 \\
\end{array}
\right)  
\right.\right.\\
&\left. \left.- G_{2,4}^{2,0}\left(\frac{m\tilde{\gamma} ^2 (\frac{1}{2}- \tilde{y})^2}{4 (k_B T+2\epsilon_F)}| 
\begin{array}{c}
	-2,1 \\
	-\frac{3}{2},-\frac{3}{2},-2,-2 \\
\end{array}
\right)\right) + \right.\nonumber \\
& \left.\left(\frac{1}{2} +\tilde{y}\right)^4 \left(G_{2,4}^{2,0}\left(-\frac{m\tilde{\gamma}^2 (\frac{1}{2}+ \tilde{y})^2}{4 (k_B T-2 \epsilon_F)}| \nonumber
\begin{array}{c}
	-2,1 \\
	-\frac{3}{2},-\frac{3}{2},-2,-2 \\
\end{array}
\right)  
\right.\right.\\
&\left. \left.- G_{2,4}^{2,0}\left(\frac{m\tilde{\gamma} ^2 (\frac{1}{2}+ \tilde{y})^2}{4 (k_B T+2\epsilon_F)}| 
\begin{array}{c}
	-2,1 \\
	-\frac{3}{2},-\frac{3}{2},-2,-2 \\
\end{array}
\right)\right) \right].\nonumber
\end{align}
If we integrate the coordinate $y$ over the width of the channel, we arrive at the average current
\begin{align}
&I= \frac{e\tilde{E}_0 W m \tilde{\gamma} }{3\pi \hbar^2} + \frac{e\tilde{E}_0 W m^2 \tilde{\gamma}^3}{64 \hbar^2 k_B T} \cdot \label{currentnonzeroT}\\
&\left[ G_{3,5}^{2,1}\left(-\frac{m \tilde{\gamma}^2}{4 (k_B T-2 \epsilon_F)}|
\begin{array}{c}
	-\frac{3}{2},-2,1 \\
	-\frac{3}{2},-\frac{3}{2},-\frac{5}{2},-2,-2 \\
\end{array}
\right) \right. \nonumber \\
&\left.-G_{3,5}^{2,1}\left(\frac{m \tilde{\gamma}^2}{4 (k_B T+2 \epsilon_F)}|
\begin{array}{c}
	-\frac{3}{2},-2,1 \\
	-\frac{3}{2},-\frac{3}{2},-\frac{5}{2},-2,-2 \\
\end{array}
\right) \right]. \nonumber
\end{align}
For low temperatures $k_B T \ll \epsilon_F$, we can expand Eq. \eqref{currentnonzeroT} up to second order in $k_B T/\epsilon_F$ yielding 
\begin{align}
    &I= \frac{e\tilde{E}_0 W m \tilde{\gamma}}{3\pi \hbar^2} + \frac{e\tilde{E}_0 W m^2 \tilde{\gamma}^3}{64 \hbar^2 \epsilon_F} \cdot \nonumber  \\
    &\left[ 
    \left. -G_{3,5}^{2,1}\left(\frac{m \tilde{\gamma}^2}{8 \epsilon_F}\left|
\begin{array}{c}
 -\frac{3}{2},-2,0 \\
 -\frac{3}{2},-\frac{3}{2},-\frac{5}{2},-2,-2 \\
\end{array}
\right.\right) \right. \right.\nonumber\\
&\left. + \left(\frac{k_B T}{2 \epsilon_F}\right)^2 F\left(\frac{m\tilde{\gamma}^2}{8\epsilon_F}\right)
\right],
\label{currentsmallT}
\end{align}
where we define the function $F\left(z\right)$ as
\begin{align}
   F\left(z\right)= &-G_{3,5}^{2,1}\left(z\left|
\begin{array}{c}
 -\frac{3}{2},-2,0 \\
 -\frac{3}{2},-\frac{3}{2},-\frac{5}{2},-2,-2 \\
\end{array}
\right.\right) \nonumber\\
&+G_{3,5}^{2,1}\left(z\left|
\begin{array}{c}
 -\frac{3}{2},0,1 \\
 -\frac{3}{2},-\frac{3}{2},-\frac{5}{2},-2,2 \\
\end{array}
\right.\right) \nonumber\\
&+\frac{1}{6}G_{3,5}^{2,1}\left(z\left|
\begin{array}{c}
 -\frac{3}{2},0,1 \\
 -\frac{3}{2},-\frac{3}{2},-\frac{5}{2},-2,3 \\
\end{array}
\right.\right).
\end{align}
We note that, in the limit $T=0$, Eq. \eqref{currentsmallT} becomes Eq. \eqref{currentsol}. As verified in Sec.~IV, \textit{cf.} Fig. \ref{subfig:d}, the analytical result for the current \eqref{currentsmallT} resembles the full numerical result for weak electron-electron scattering rates ($\tilde{\gamma}_{ee} \ll \tilde{\gamma}_d$).\\
In general, electron-electron scattering is temperature dependent, while impurity scattering does not depend on temperature. To mimic this behavior, we use the following ansatz for the total scattering rate
\begin{equation}
    \tilde{\gamma} = \tilde{\gamma}_d + \tilde{\gamma}_{ee} = \tilde{\gamma}_d + C(k_B T)^\alpha,
    \label{gamma_Talpha}
\end{equation}
where we assume $\tilde{\gamma}_d \gg C(k_B T)^\alpha$ to stay in the validity regime of our analytical approach. We keep the exponent $\alpha$ as a variable to be able to follow it in the expression below. In general, $\alpha  = 2$.
Inserting this ansatz into Eq. \eqref{currentsmallT}, we obtain in leading order of $C(k_B T)^\alpha/\tilde{\gamma}_d$
\begin{align}
        &I= \frac{e\tilde{E}_0 W m \tilde{\gamma}_d}{3\pi \hbar^2} + \frac{e\tilde{E}_0 W m^2 \tilde{\gamma}_d^3}{64 \hbar^2 \epsilon_F} \cdot \nonumber  \\
    &\left[ 
    \left. -G_{3,5}^{2,1}\left(\frac{m \tilde{\gamma}^2}{8 \epsilon_F}\left|
\begin{array}{c}
 -\frac{3}{2},-2,0 \\
 -\frac{3}{2},-\frac{3}{2},-\frac{5}{2},-2,-2 \\
\end{array}
\right.\right) \right. \right.\nonumber\\
&\left. + \left(\frac{k_B T}{2 \epsilon_F}\right)^2 F\left(\frac{m\tilde{\gamma}^2}{8\epsilon_F}\right)
\right] +\nonumber\\
&+\frac{e\tilde{E}_0 W m C(k_B T)^\alpha}{3\pi \hbar^2} + \frac{e\tilde{E}_0 W m^2 3\tilde{\gamma}_d^2C(k_B T)^\alpha}{64 \hbar^2 \epsilon_F} \cdot \nonumber  \\
    &\left[ 
    \left. -G_{3,5}^{2,1}\left(\frac{m \tilde{\gamma}^2}{8 \epsilon_F}\left|
\begin{array}{c}
 -\frac{3}{2},-2,0 \\
 -\frac{3}{2},-\frac{3}{2},-\frac{5}{2},-2,-2 \\
\end{array}
\right.\right) \right. \right.\nonumber\\
&\left. + \left(\frac{k_B T}{2 \epsilon_F}\right)^2 F\left(\frac{m\tilde{\gamma}^2}{8\epsilon_F}\right)
\right] + \nonumber\\
&+ \frac{e\tilde{E}_0 W m^2 2\tilde{\gamma}_d^2C(k_B T)^\alpha}{64 \hbar^2 \epsilon_F} \cdot \nonumber  \\
    &\left[ 
    \left. -G_{3,5}^{2,1}\left(\frac{m \tilde{\gamma}^2}{8 \epsilon_F}\left|
\begin{array}{c}
 -\frac{3}{2},0,0 \\
 -\frac{3}{2},-\frac{3}{2},-\frac{5}{2},-2,1 \\
\end{array}
\right.\right) \right. \right.\nonumber\\
&\left. + \left(\frac{k_B T}{2 \epsilon_F}\right)^2 H\left(\frac{m\tilde{\gamma}^2}{8\epsilon_F}\right)
\right],
    \label{currentsmalleescattering}
\end{align}
where $H(z)$ is defined as
\begin{align}
    H\left(z\right)= &-G_{3,5}^{2,1}\left(z\left|
\begin{array}{c}
 -\frac{3}{2},0,0 \\
 -\frac{3}{2},-\frac{3}{2},-\frac{5}{2},-2,1 \\
\end{array}
\right.\right) \nonumber\\
&+G_{4,6}^{2,2}\left(z\left|
\begin{array}{c}
 -\frac{3}{2},0,0,1 \\
 -\frac{3}{2},-\frac{3}{2},-\frac{5}{2},-2,1,2 \\
\end{array}
\right.\right) \nonumber\\
&+\frac{1}{6}G_{4,6}^{2,2}\left(z\left|
\begin{array}{c}
 -\frac{3}{2},0,0,1 \\
 -\frac{3}{2},-\frac{3}{2},-\frac{5}{2},-2,1,3 \\
\end{array}
\right.\right).  
\end{align}
We emphasize that Eq. \eqref{currentsmalleescattering} describes analytically the full crossover of the average current from ballistic to diffusive transport at finite temperatures in presence of weak electron-electron scattering in a channel geometry with diffusive boundary conditions. In principle, it can be employed to determine the exponent $\alpha$. For small temperatures, $\alpha$ should be equal to 2 in 2D electron transport, even if we take into account a mode-dependent scattering rate \cite{Kryhin2023, Hofmann2023}. The reason is that the electron-electron scattering rate of the long-lived modes at small temperatures is predicted to scale as $T^2$ with temperature $T$. Those modes should be the most relevant ones in the validity regime of Eq. \eqref{currentsmalleescattering}. i.e. weak electron-electron scattering.

\section{Conclusions}
\label{conclusion}
We derive an analytical solution for the current density and average current at zero temperature that describes the transition from ballistic to diffusive transport. Additionally, we analyze the transition between different scattering regimes (ballistic, diffusive, hydrodynamic) numerically. We illustrate these transitions by characteristic current density profiles. They imply that it is difficult to distinguish ballistic from hydrodynamic transport based on current density profiles. In the regime of weak electron-electron scattering as compared to impurity scattering, we derive an analytical expression for the temperature dependence of the average current in channel geometry with diffusive boundaries.

\begin{acknowledgments}
We thank V.~Kornich and P.~O.~Sukhachov for stimulating discussions.
This work was supported by the W\"urzburg-Dresden Cluster of Excellence ct.qmat, EXC2147, project-id 390858490, the DFG (SFB 1170), and the Bavarian Ministry of Economic Affairs, Regional Development and Energy within the High-Tech Agenda Project “Bausteine f\"ur das Quanten Computing auf Basis topologischer Materialen”.
P.~P. acknowledges funding by the Free State of Bavaria for the Institute for Topological Insulators.
\end{acknowledgments}

\bibliography{Ref.bib}
\onecolumngrid
\newpage
\setcounter{secnumdepth}{2}
\appendix
\renewcommand{\thesection}{\Alph{section}}
\renewcommand{\thesubsection}{\Alph{section}\arabic{subsection}}
\renewcommand{\thefigure}{\Alph{section}\arabic{figure}}
\renewcommand{\theequation}{\Alph{section}\arabic{subsection}.\arabic{equation}}
\numberwithin{equation}{section}
\renewcommand{\thetable}{\Alph{section}\arabic{subsection}.\arabic{table}}
\numberwithin{table}{subsection}
\makeatletter
\renewcommand{\p@subsection}{}
\makeatother
\end{document}